\documentclass[onecolumn,showpacs,showkeys,preprintnumbers,amsmath,amssymb,groupaddress]{revtex4}
\usepackage{mathbbold}

\usepackage{graphicx}
\usepackage{dcolumn}
\usepackage{bm}

\begin{document}

\title{The scalable quantum computation based on quantum dot systems}
\author{Jian-Qi Zhang}
\author{Ya-Fei Yu}
\author{Xun-Li Feng}
\author{Zhi-Ming Zhang}
\email[Corresponding author Email: ]{zmzhang@scnu.edu.cn}
\affiliation{Key Laboratory of Photonic Information Technology of Guangdong Higher
Education Institutes, SIPSE $\&$ LQIT, South China Normal University,
Guangzhou 510006, China\\
}

\begin{abstract}
We propose a scheme for realizing the scalable quantum computation
based on nonidentical quantum dots trapped in a single-mode
waveguide. In this system, the quantum dots simultaneously interact
with a large detuned waveguide and classical light fields. During
the process, neither the waveguide mode nor the quantum dots are
excited, while the sub-system composed of any two quantum dots can
acquire phases conditional upon the states of these two quantum dots
and the certain detunings between the waveguide mode and
corresponding external light fields. Therefore, it can be used to
realize selective quantum phase gates, graph states, $N$-qubit
controlled phase $\pi $ gates, and cluster states.
\end{abstract}

\date{\today }
\maketitle

\section{Introduction}

Semiconductor quantum dots (QDs) embedded in a photonic crystal (PC)
cavity provides a promising system to investigate cavity quantum
electrodynamics and quantum information processing (QIP) in the
solid state \cite{01}. In the past decade, it has attracted
considerable experimental and theoretical attention. Both the weak
and strong couplings have been achieved in experiment
\cite{02,0201,0202,0203,0204,0205}. However, the practical and
useful QIP requires a large number of qubits, and it is difficult to
achieve so many spatial separation QDs in a PC cavity experimentally
\cite{03}. On the other hand, experiments have proved that the above
system also can be used to harvest single photons by coupling a
single QD to an enhanced cavity mode\cite{0205}. Nevertheless, since
generated single photons must be coupled out of the cavity, the
overall efficiency of this kind of single-photon source isn't high
enough. In order to overcome this challenge, Hughes \textit{et al.}
presented several theory proposals based on PC waveguides
\cite{04,0401}. These schemes show single QDs also can coupled a PC
waveguides efficiently. And it has been proved in experiment by
Lund-Hansen \textit{et al.} \cite{prl101}.

Very recently, Feng \textit{et al.} proposed a scheme to realize a
quantum computation with atoms in decoherence-free subspace by using
a dispersive atom-cavity interaction driven by strong classical
laser fields \cite{24}. But their proposal is based on identical
qubits, and each qubit is driven with four laser fields. Motivated
by these works, we present a scheme for realizing the scalable
quantum computation based on nonidentical QDs trapped in a
single-mode PC waveguide. In this scheme, any two QDs can acquire
different phases conditional upon their different states and
corresponding detunings between the waveguide mode and external
light fields. And selective gate operations for any two QDs can be
acquire in this way. For this reason, this scheme also can be
employed to achieve graph states, $N$-qubit controlled phase $\pi $
gate (NCZ gate), and cluster states with different number of gate
operations. During the gate operation, neither the QDs nor the
waveguide is excited. Comparing with Ref. \cite{24}, the logical
gate is extended to nonidentical qubits, and the number of laser
fields is decreased. In addition, this scheme is the first scheme to
realize the scalable quantum computation with spatially separated
and nonidentical QDs.

The organization of this paper is as follows. In Sec.\ref{seck}, we
introduce the theoretical model and effective Hamiltonian. In Sec.\ref{sec31}%
, we present how to realize the selective quantum phase gate between
any two QDs. In Sec.\ref{sec4}, we give the operations to achieve
the graph states, NCZ gate, and cluster state. In Sec.\ref{sec5}, we
show the simulation and realizability of the above gate operations
and entangled states. The conclusion is given in Sec.\ref{sec6}.

\section{Theoretical Model and effective Hamiltonian}

\label{seck}

Let us consider that $N$ charged and spatially separated GaAs/AlGaAs QDs are
trapped in a single-mode waveguide. Each dot has two lower states $|g\rangle
=|\uparrow \rangle $, $|f\rangle =|\downarrow \rangle $ and a higher state $%
|e\rangle =|\uparrow \downarrow \Uparrow \rangle $, here ($|\uparrow \rangle
$, $|\downarrow \rangle $) and ($|\Uparrow \rangle $, $|\Downarrow \rangle $%
) denote the spin up and spin down for electron and hole, respectively. The
transitions $|g\rangle \leftrightarrow |e\rangle $\ and $|f\rangle
\leftrightarrow |e\rangle $ are correspondingly coupled to the vertical
polarization and horizontal polarization lights \cite{17,35}. With the
choice of the fields with the vertical polarization, the state $|f\rangle $
can be treated as an auxiliary state, while the transition $|g\rangle $ $%
\leftrightarrow $ $|e\rangle $ can coupled to the waveguide mode and
classical laser fields. In this situation, the Hamiltonian describing the
interaction between QDs and fields can be written as:

\begin{equation}
\begin{array}{rcl}
\hat{H}_{I} & = & \sum\limits_{j=1}^{N}(g_{j}ae^{i\Delta _{j}^{C}t}+\dfrac{%
\Omega _{j}}{2}e^{i\Delta _{j}t}+\dfrac{\Omega _{j}^{^{\prime }}}{2}%
e^{-i\Delta _{j}^{^{\prime }}t})\sigma _{j}^{+}+H.c.,%
\end{array}
\label{eq1}
\end{equation}%
where$\ g_{j}$ is the coupling constant between QD $j$ and the waveguide
mode with the detuning $\Delta _{j}^{C}$, $a$ is the annihilation operator
for the waveguide mode, $\Omega _{j}$ and $\Omega _{j}^{^{\prime }}$ are the
Rabi frequencies driven by the laser fields with the detunings $\Delta
_{j}^{{}}$ and $\Delta _{j}^{^{\prime }}$, respectively, and $\sigma
_{j}^{+}=|e\rangle _{j}\langle g|$ (see FIG.1 ).

\begin{figure}[tbph]
\includegraphics[width=6cm]{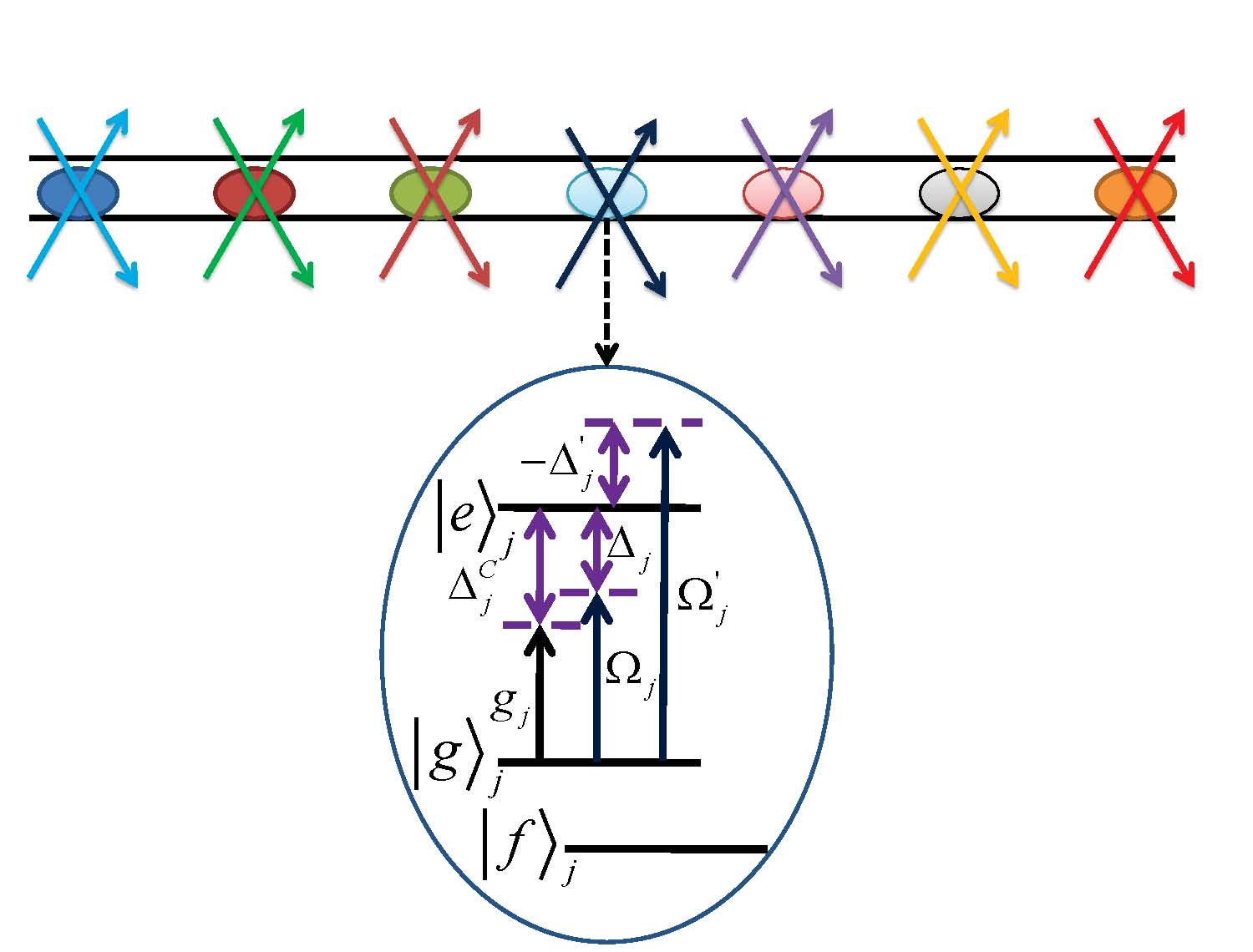}
\caption{(Color online) Each of QDs is driven with two classical fields and
one quantum field.}
\label{fig 1}
\end{figure}

In order to derive the effective Hamiltonian of the system, we will use the
method proposed in Refs. \cite{24,28,2801} under the following conditions: (1) $%
|\Omega _{j}|=|\Omega _{j}^{^{\prime }}|$; (2) $\Delta _{j}=\Delta
_{j}^{^{\prime }}$; (3) the large detuning condition: $|\Delta
_{j}|,|\Delta _{j}^{^{\prime }}|\gg |g_{j}|,|\Omega _{j}|,|\Omega
_{j}^{^{\prime }}|$; (4) $\delta _{j}=\Delta _{j}^{C}-\Delta
_{j}^{{}}$. The first condition together with the second condition
can cancel the Stark shifts and related terms completely. Under the
large detuning condition, the probability for QDs absorbing photons
from the light fields or being excited is negligible. The last
condition ensures that $\delta _{j}$ is only related with detuning
between the waveguide field and light fields. In this situation, if
the QDs are initial in the ground states, the excited states will
not be populated and can be adiabatically eliminated. Thus we can
obtain the effective Hamiltonian:

\begin{equation}
\hat{H}_{eff-1}=-\sum_{j=1}^{N}(\dfrac{|g_{j}|^{2}}{\Delta _{j}^{C}}%
a^{+}a+\lambda _{j}ae^{i\delta _{j}t}+\lambda _{j}^{\ast }a^{+}e^{-i\delta
_{j}t})|g\rangle _{j}\langle g|,  \label{eq2}
\end{equation}%
where $\lambda _{j}=\dfrac{\Omega _{j}^{\ast }g_{j}}{4}(\dfrac{1}{\Delta
_{j}^{{}}}+\dfrac{1}{\Delta _{j}^{^{C}}})$. The first term describes a Stark
shift caused by the waveguide mode, the last two terms shows the indirect
coupling between laser fields and waveguide field, which is caused by the
virtually excited QDs.

Under the condition $\delta _{j}\gg |g_{j}|^{2}/\Delta _{j}^{C} ,|\lambda
_{j}|$, the waveguide mode cannot exchange energy with the classical fields,
the photon in the waveguide is only virtually excited, and any two QDs can
interfere with each other. So the effective Hamiltonian takes the form:
\begin{widetext}
\begin{equation}
\begin{array}{rcl}
\hat{H}_{eff-2} & = & \sum\limits_{j=1}^{N}(-\dfrac{|g_{j}|^{2}}{\Delta
_{j}^{C}}a^{+}a+\eta _{jj})\sigma _{j}^{-}\sigma
_{j}^{+}+2\sum\limits_{j=1,k=1,j\neq k}^{N}\eta _{jk}\sigma _{j}^{-}\sigma
_{j}^{+}\sigma _{k}^{-}\sigma _{k}^{+}\cos (\delta _{jk}t)%
\end{array}%
,
\end{equation}%
\end{widetext}
where $\delta _{jk}=\delta _{j}-\delta _{k}$ and $\eta
_{jk}=\dfrac{|\lambda _{j}\lambda _{k}|}{2}(\dfrac{1}{\delta
_{j}}+\dfrac{1}{\delta _{k}})$. With the initial state for the
waveguide mode being in the vacuum state assumed, the effective
Hamiltonian reduces to
\begin{widetext}
\begin{equation}
\begin{array}{rcl}
\hat{H}_{eff} & = & \left\{
\begin{array}{cl}
\sum\limits_{j=1}^{N}\eta _{jj}\sigma _{j}^{-}\sigma
_{j}^{+}+2\sum\limits_{j=1,k=1,j\neq k}^{N}\eta _{jk}\sigma _{j}^{-}\sigma
_{j}^{+}\sigma _{k}^{-}\sigma _{k}^{+}, & \delta _{j}=\delta _{k} \\
\sum\limits_{j=1}^{N}\eta _{jj}\sigma _{j}^{-}\sigma
_{j}^{+}+2\sum\limits_{j=1,k=1,j\neq k}^{N}\eta _{jk}\sigma _{j}^{-}\sigma
_{j}^{+}\sigma _{k}^{-}\sigma _{k}^{+}\cos (\delta _{jk}t) & \delta _{j}\neq
\delta _{k}%
\end{array}%
\right.%
\end{array}
\label{eq302}
\end{equation}
\end{widetext}
This equation can be understood as follows. Under the condition of $\delta
_{j}=\delta _{k}$, with the laser field acting, QDs will take place the
Stark shifts and acquire the virtual excitation, and this virtual excitation
will induce the coupling between the vacuum waveguide mode and classical
fields. As the Stark shifts are nonlinear in the number of any two QDs in
the state $|g\rangle $, the subsystem composed by arbitrary two QDs can
acquire a phase conditional upon the number of these two QDs in the state $%
|g\rangle $. On the contrary, in the case of $\delta _{j}\neq \delta _{k}$,
there might be not additional phase for the two QDs in the state $|g\rangle $%
. As a result, this system can be employed to construct the selective
controlled phase gate.

\section{The selective quantum controlled phase}

\label{sec31}

Now, we take QDs $m$ and $n$ as an example to discuss how to
construct the selective quantum controlled phase gates with
arbitrary two QDs. In order to do so, states $|f\rangle $ and
$|g\rangle $ are used to store the quantum information at first. In
the case of $\delta _{m}=\delta _{n}$, by using the appropriate
light fields for QDs $m$ and $n$, we can get
$\delta _{m}=\delta _{n}=n\delta _{0}$, $\lambda _{m}=\lambda _{n}=\sqrt{n}%
\lambda _{0}$. Then, the effective Hamiltonian (\ref{eq302}) takes the form
of%
\begin{equation}
\begin{array}{rcl}
\hat{H}_{es} & = & \epsilon (\sum\limits_{j=m,n}\sigma _{j}^{-}\sigma
_{j}^{+}+2\sigma _{m}^{-}\sigma _{m}^{+}\sigma _{n}^{-}\sigma _{n}^{+}),%
\end{array}
\label{eq411}
\end{equation}%
here, $\epsilon =\dfrac{|\lambda _{0}|^{2}}{2\delta _{0}}$. After that,
according to the Hamiltonian (\ref{eq411}) for the same detuning, the
evolutions of the logical states are:

\begin{equation}
\left\{
\begin{array}{rcl}
|ff\rangle _{m,n} & \rightarrow & |ff\rangle _{m,n}, \\
|fg\rangle _{m,n} & \rightarrow & \exp (-i\epsilon t)|fg\rangle _{m,n}, \\
|gf\rangle _{m,n} & \rightarrow & \exp (-i\epsilon t)|gf\rangle _{m,n}, \\
|gg\rangle _{m,n} & \rightarrow & \exp (-i4\epsilon t)|gg\rangle _{m,n}.%
\end{array}%
\right.  \label{eq001}
\end{equation}%
With the application of the single-qubit operation $|g\rangle
_{j}\rightarrow \exp (i\epsilon t)|g\rangle _{j}$, the above equation can be
rewritten as%
\begin{equation}
\left\{
\begin{array}{rcl}
|ff\rangle _{m,n} & \rightarrow & |ff\rangle _{m,n}, \\
|fg\rangle _{m,n} & \rightarrow & |fg\rangle _{m,n}, \\
|gf\rangle _{m,n} & \rightarrow & |gf\rangle _{m,n}, \\
|gg\rangle _{m,n} & \rightarrow & \exp (-2i\epsilon t)|gg\rangle _{m,n}.%
\end{array}%
\right.  \label{eq002}
\end{equation}%
This transformation for QDs $m$ and $n$ corresponds to the quantum
phase gate operation, in which if and only if both controlling and
controlled qubits are in the states $|g\rangle $, there will be an
additional phase in the system. During the operation, none of QDs
and waveguide modes is excited. It is worth to point out that,
although QDs are nonidentical, $\delta _{j}=\Delta _{j}^{C}-\Delta
_{j}$ is a tunable constant, which is decided by the frequency of
detuning between the laser field and waveguide mode. Therefore, this
system can construct the controlled phase gate with different QDs.

On the other hand, in the case of $\delta _{m}\neq \delta _{n}$, with the
choice of appropriate light fields for QDs $m$ and $n$, we can get $\delta
_{j}=j\delta _{0}$, $\lambda _{j}=\sqrt{j}\lambda _{0}$, $\eta
_{jj}=\epsilon $, and $\delta _{mn}=(m-n)\delta _{0}$. So the effective
Hamiltonian (\ref{eq302}) takes the form:%
\begin{equation}
\begin{array}{rcl}
\hat{H}_{ed} & = & \sum\limits_{j=m,n}\epsilon \sigma _{j}^{-}\sigma
_{j}^{+}+2\eta _{mn}\sigma _{m}^{-}\sigma _{m}^{+}\sigma _{n}^{-}\sigma
_{n}^{+}\cos (\delta _{mn}t).%
\end{array}
\label{eq412}
\end{equation}%
And the time evolutions of four logical states for the two QDs $m$
and $n$, under the Hamiltonian (\ref{eq412}) for the different
detunings, are given by:

\begin{equation}
\left\{
\begin{array}{rcl}
|ff\rangle _{m,n} & \rightarrow & |ff\rangle _{m,n}, \\
|fg\rangle _{m,n} & \rightarrow & \exp (-i\epsilon t)|fg\rangle _{m,n}, \\
|gf\rangle _{m,n} & \rightarrow & \exp (-i\epsilon t)|gf\rangle _{m,n}, \\
|gg\rangle _{m,n} & \rightarrow & \exp (-i2(\epsilon t+\dfrac{\eta _{mn}}{%
\delta _{mn}}\sin (\delta _{mn}t))|gg\rangle _{m,n}.%
\end{array}%
\right.  \label{eq003}
\end{equation}%
After the performance of the single-qubit operation $|g\rangle
_{j}\rightarrow \exp (i\epsilon t)|g\rangle _{j}$, there is%
\begin{equation}
\left\{
\begin{array}{rcl}
|ff\rangle _{m,n} & \rightarrow & |ff\rangle _{m,n}, \\
|fg\rangle _{m,n} & \rightarrow & |fg\rangle _{m,n}, \\
|gf\rangle _{m,n} & \rightarrow & |gf\rangle _{m,n}, \\
|gg\rangle _{m,n} & \rightarrow & \exp (-i2\dfrac{\eta _{mn}}{\delta _{mn}}%
\sin ((m-n)\delta _{0}t))|gg\rangle _{m,n}.%
\end{array}%
\right.  \label{eq004}
\end{equation}%
It means, in the situation of $\delta _{0}t=k\pi $ for $k=1,2,3,....$, there
will be no the controlled phase gate operation for the QDs $m$ and $n$.

As a result, with the choice of $2\epsilon t=\pi $, this system can
realize the several selective controlled phase $\pi$ gate (SCZ gate)
operations for the different QDs with different detuning at the same
time.

\section{Graph states, NCZ gate, and cluster states}

\label{sec4}

Here, we show how to acquire graph states, NCZ gate, and cluster
states in the system. At first, we will review the definition of an
N-qubit graph
state in brief. In a system of $N$ qubits, if each qubit is in the state $%
|+\rangle =(|g\rangle +|f\rangle )/\sqrt{2}$, and to all pairs
$\{m,n\}$ of qubits joined by a controlled phase $\pi $ gate
($CZ_{m,n}$ gate), these $N$ qubits are in the graph state
\cite{pra}, which can be expressed as

\begin{equation}
\begin{array}{lll}
|G\rangle & = & \otimes _{m,n\in N}CZ_{m,n}(\otimes _{j\in N}|+\rangle _{j})%
\end{array}%
.  \label{eq41}
\end{equation}

We will prepare this state as follows. Assume $N+1$ QDs are in the
initial state $|\Psi \rangle _{N+1}=\otimes _{j\in N+1}|+\rangle
_{j}$, they are trapped in a single-mode waveguide, and
simultaneously driven by the
appropriate laser fields. The transition $|g\rangle $ $\leftrightarrow $ $%
|e\rangle $ is initial far off resonant with the fields, and all the
detunings between the single-mode waveguide and laser fields are the
same. If the waveguide mode is initial in the vacuum state, the
state evolution for any two QDs is governed by Eq.(\ref{eq002}).
Waiting for a controlled phase $\pi $ gate operation time, a graph
state for $N+1$ QDs can be created. After the above graph state is
generated, with the implement of controlled phase $\pi $ gate
operations for $N$ of these $N+1$ QDs again, a NCZ gate can be
constructed. The process of generating the graph state and NCZ gate
can refer the figures in Ref. \cite{lin}. In addition, as the
selective controlled phase $\pi $\ gate can be realized in several
different groups at the same time, with the choice of $\delta
_{J}=J\delta _{0}$ and $\lambda _{J}=\sqrt{J}\lambda _{0}$ for group
$J$, several different graph states or NCZ gates can be achieved
simultaneously.

\begin{figure}[tbph]
\includegraphics[width=6cm]{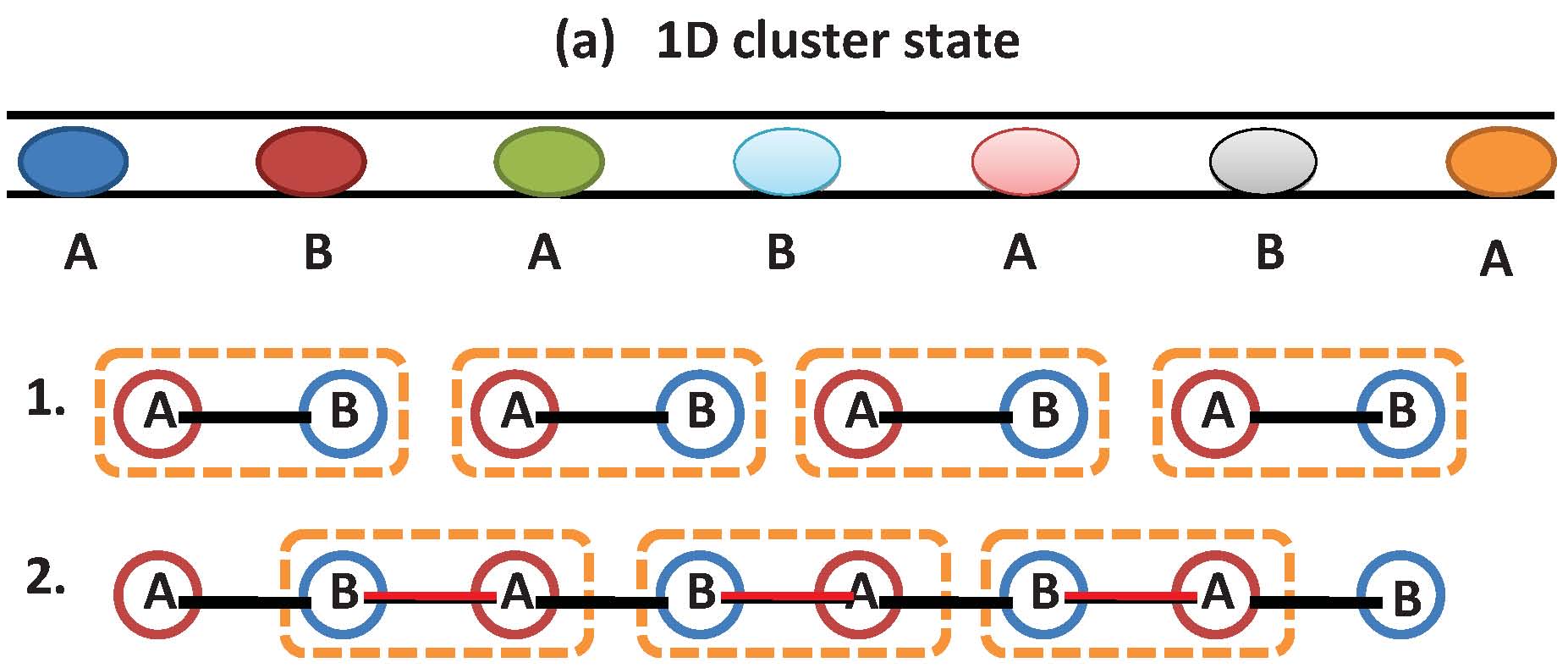} \includegraphics[width=6cm]{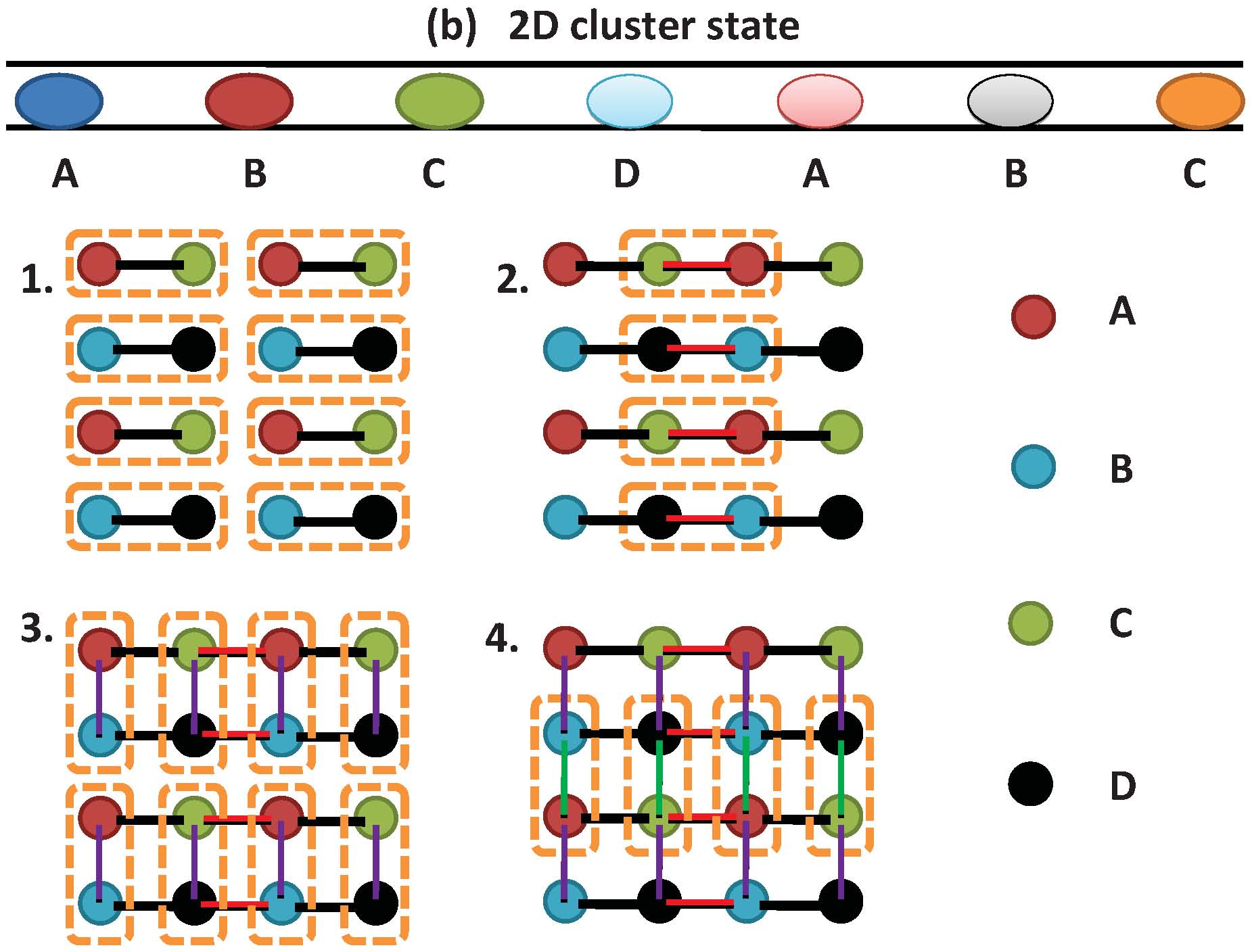}
\caption{(Color online) The operations for generate cluster states}
\label{cacb}
\end{figure}

As it is well known, cluster states can be acquired by the local
unitary operation from graph states. If each dot is initial in the state $%
|+\rangle $ and encoded with ABABABAB..., a 1D cluster state can be
realized by two steps (see FIG.\ref{cacb}-(a)). First step: Apply
the SCZ gate operations for A and B; second step:  Apply the SCZ
gate operations for B and A. After that, a 1D cluster state is
achieved. The operations for constructing the 2D cluster states are
listed as follows (see FIG.\ref{cacb}-(b)). i) Name dots with
ABCDABCDABCD...; ii) With the application of SCZ gate operations for
A and B, C and D, successively, several 1D cluster states are
generated ; iii) After applying the SCZ gate operations for A and C,
B and D, respectively, a 2D cluster state is created.

\section{Simulations of decoherence}

\label{sec5} 
\begin{figure}[tbph]
\includegraphics[width=8cm]{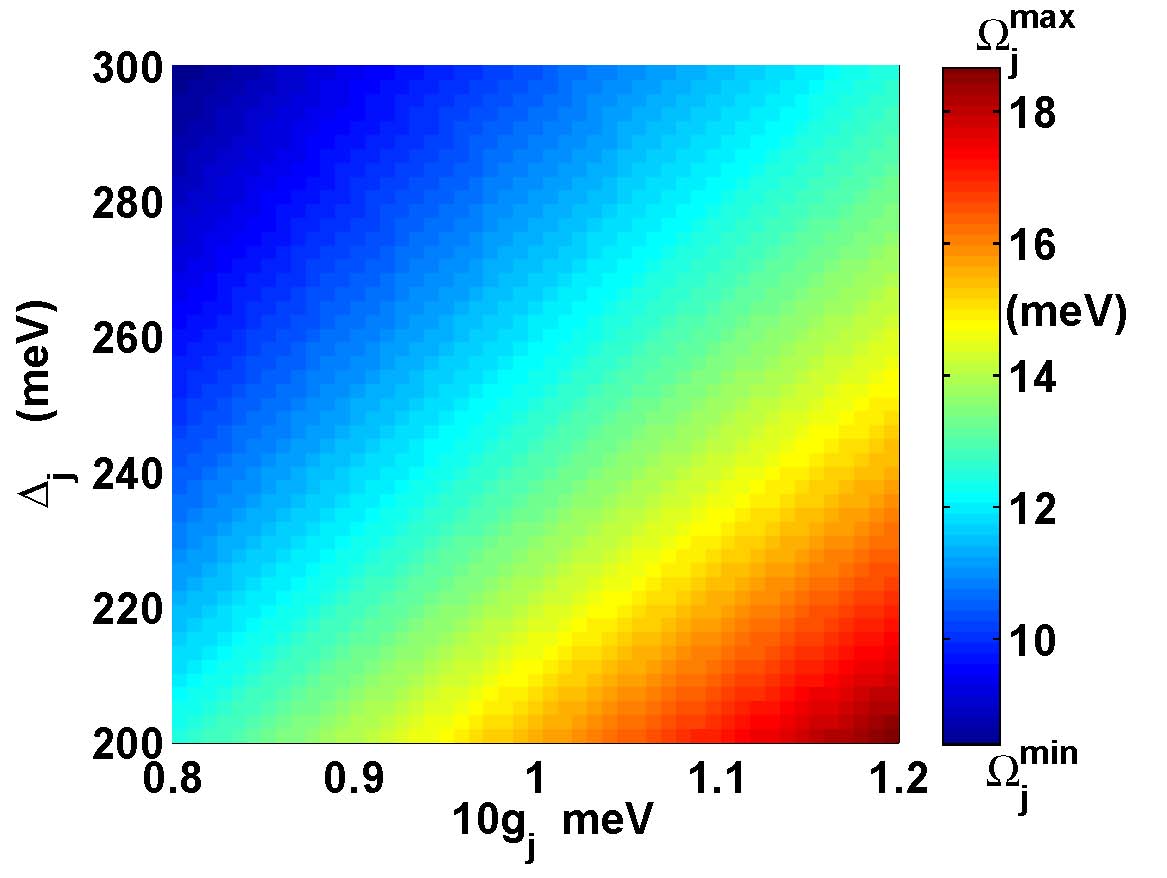}
\caption{(Color online) $\Omega_{j}$ versus $g_{j}$ and $\Delta_{j}$}
\label{fig1}
\end{figure}

\begin{figure}[tbph]
\includegraphics[width=4cm]{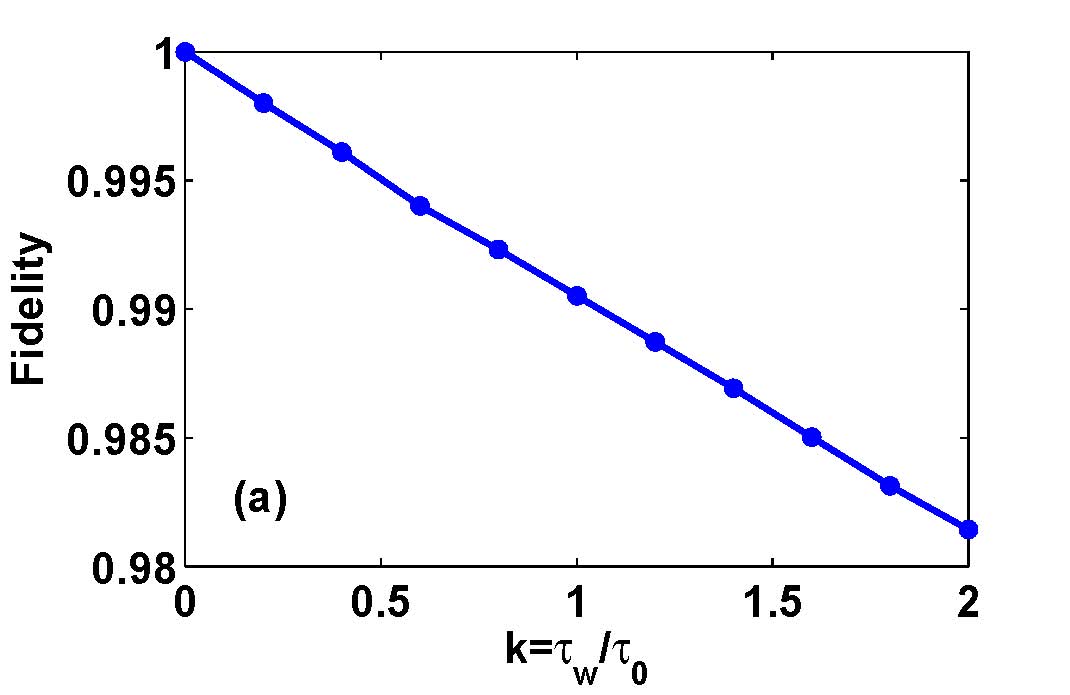} \includegraphics[width=4cm]{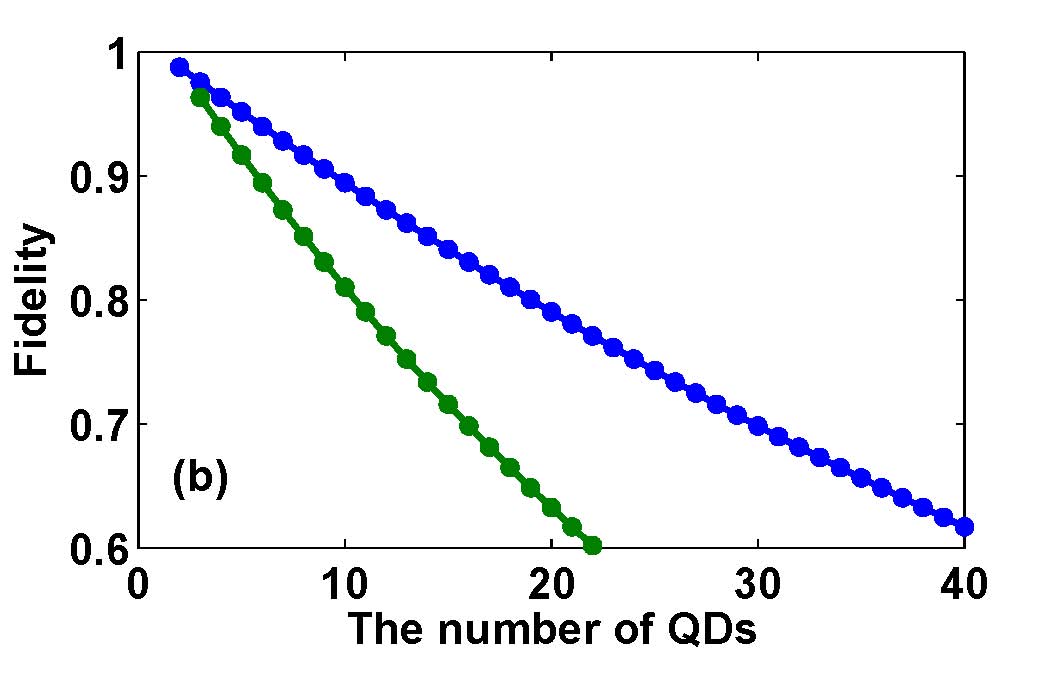}
\caption{(Color online) (a) Numerical simulation of the fidelity of
the any two qubits CZ$_{A,B}$
gates versus the waveguide decay time $\tau_{0}$, with the parameters $g_{A}=0.10meV$, $%
g_{B}=0.08meV$, $\Omega _{A}=10meV$ $\Omega _{B}=13.75meV$. The
detunings of blue line are given by $\Delta _{A}=200.00meV$, $\Delta
_{B}=220.00meV$, and the detunings of green line are given by
$\Delta _{A}=200.09meV$, $\Delta _{B}=220.09meV$, respectively.
$\tau_{0}$ is the decay time. (b) the fidelity of the graph states
and NCZ gates versus the number of QDs with the decay time of
$\protect\tau_{w}$. And the Fidelities for graph states and NCZ
gates are the blue line and green line, respectively.}
\label{fidelity}
\end{figure}

\begin{figure}[tbph]
\includegraphics[width=6cm]{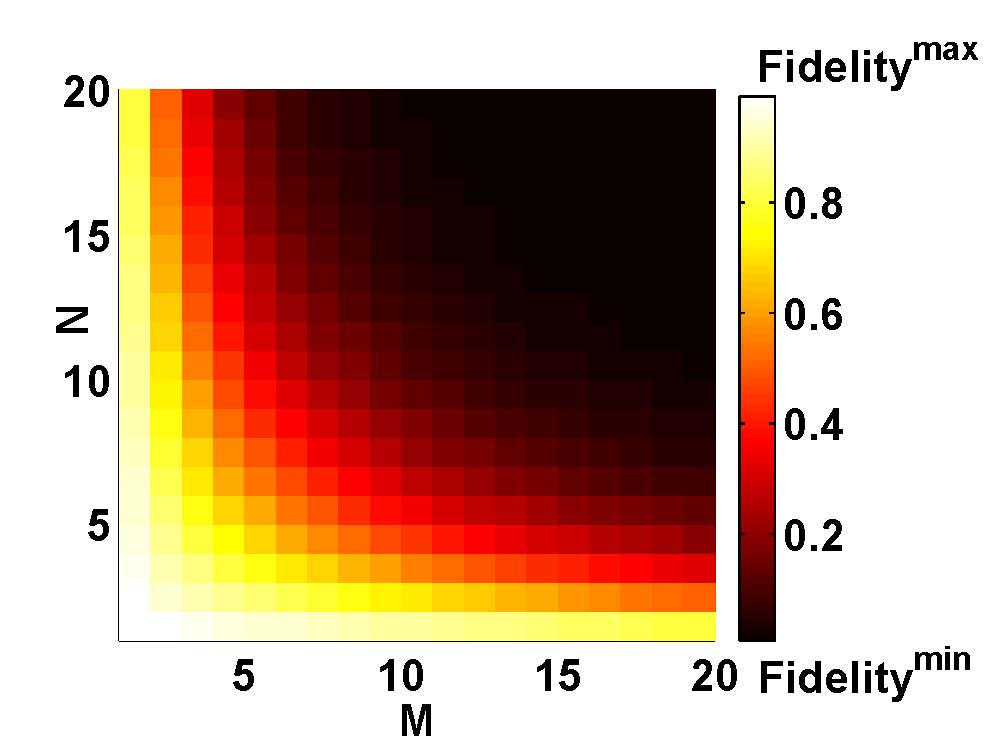}
\caption{(Color online) Numerical simulation of the fidelity of the cluster
states versus the number of $M\times N$ QDs with the decay time of $\protect%
\tau _{w}$}
\label{cluster}
\end{figure}

In the following, let us discuss the realizability of
the experiment. According to the above discussion, the influences of
spontaneous emission from the excited states and the waveguide decay
can be ignored. As a matter of fact, under the condition of the
largely detuned couplings the excited state is rarely populated, so
the influence of the spontaneous emission can be neglected, and the
main decoherence effect in our scheme is due to waveguide decay.
Then, the master equation can be given as follows:
\begin{equation}
\begin{array}{rcl}
\dot{\rho} & = & -i[H_{I},\rho ]+\frac{\gamma }{2}(2a\rho a^{+}-a^{+}a\rho
-\rho a^{+}a)%
\end{array}
\label{eq02221}
\end{equation}%
where $\gamma=1/\tau_{w}$ is the waveguide decay rate, and
$\tau_{w}$ is the decay time of the waveguide mode. And the fidelity
of the entangled states and gate operations can be expressed as
$F=Tr(\rho \rho ^{^{\prime }})$ with $\rho (\rho ^{^{\prime }})$
being the density operator of the system in the case with (without)
waveguide decay. According to experimentally achievable parameters
in the system of QDs embedded in a single-mode waveguide
\cite{prl101,apl2010,apl201097}, the coupling of QDs and waveguide
is about $0.1meV$, the decay time for waveguide is $\tau _{w}\sim
1ns$. With the choices of the coupling constants and detunings as
FIG.\ref{fig1}, which apparently satisfy the approximation
conditions mentioned above, we can get $\lambda _{j}=0.0025meV$. The
performance of the any two QDs $A$ and $B$ controlled phase $\pi$
gate (CZ$_{A,B}$ gate) operations versus the waveguide decay time
$\tau_{0}$, and the fidelities of the entangled states and NCZ gate
operations versus the number of qubits are given in FIG.
\ref{fidelity} and FIG. \ref{cluster}.

FIG.\ref{fidelity} (a) presents, with the increase of $\tau_{w}
/\tau_{0}$, the fidelity for the two-qubit quantum controlled phase
$\pi $ gate is decreasing. It means that the waveguide decay affects
the fidelity of the gate operation largely \cite{24}. And the
fidelity is 0.9877 for $\tau_{0}=\tau _{w}=1ns$. Moreover, in this
case the gate operation time is about $50ns$, comparing with
effective decay time of waveguide $1.5\times 10^{4}ns$ ($\simeq \tau _{w}/(%
\frac{|\lambda |^{2}}{\delta ^{2}})$), it is possible to perform
hundreds two-qubit controlled phase $\pi $ gates within the
effective decay time. FIG. \ref{fidelity} and FIG. \ref{cluster}
show, with increasing the number of QDs, the fidelities for the
graph state, NCZ gate operation and cluster state decreases. It is
due to that, with the number of QDs increasing, the probability of
waveguide mode in the excited state increases. Moreover, FIG.
\ref{cluster} also presents the relationships $F_{M\times
N}=F_{N\times M}$ (for $M\times N$ QDs) and $F_{1\times
12}>F_{2\times 6}>F_{3\times 4}$. The reason for these is, the
operation time for $M\times N$ and $N\times M$ is the same, while
the operation time increases from $1\times 12$ to $3\times 4$. For
the same reason, with the same number of QDs, the fidelity for 1D
cluster state is higher than the one for 2D cluster state, which can
be seen from FIG. \ref{cluster}.

\section{Conclusion}

\label{sec6} In summary, we have shown that in a single-mode PC waveguide, $%
N $ nonidentical and spatially separated QDs can be used to realize
the scalable quantum computation with the application of the
classical light fields. During the process, neither the waveguide
mode nor the QDs are excited. The distinct advantages of the
proposed scheme are as follows: firstly, this system is scalable and
controllable; secondly, there is no waveguide photon population
involved and the QDs are almost in their ground states; thirdly, as
the QDs are non-identical, it is more practical. Therefore, we could
use this scheme to construct a kind of scalable and controllable
solid-state optical logical devices. In addition, this method opens
up a prospect to realize a scalable quantum computation in QD
system.

\section*{Acknowledgments}

This work was supported by the National Natural Science Foundation of China
(Grant No. 60978009) and the National Basic Research Program of China (Grant
Nos. 2009CB929604 and 2007CB925204).

\end{document}